\documentclass{ws-procs975x65}
\usepackage{graphicx}

\def\beq{\begin{equation}}
\def\eeq{\end{equation}}

\begin{document}

\title{Evolution of spectral and temporal properties of MAXI J1836-194 during 2011 outburst}

\author{ARGHAJIT JANA$^{1}$, DIPAK DEBNATH$^{1*}$, SANDIP K. CHAKRABARTI$^{2,1}$, SANTANU MONDAL$^{1}$, ASLAM ALI MOLLA$^{1}$, DEBJIT CHATTERJEE$^{1}$}

\address{1. Indian Centre For Space Physics, 43 Chalantika, Garia Station Road, Kolkata, 700084, India\\
2. S.N. Bose National Center for Basic Sciences, JD-Block, Salt Lake, Kolkata, 700098, India\\
$^*$E-mail: dipak@csp.res.in}

%
%

\begin{abstract}
We study transient Galatic black hole candidate MAXI~J1836-194 during its 2011 outburst using
RXTE/PCA archival data. 2.5-25~keV spectra are fitted with Two Component Advective Flow (TCAF) 
model fits file as an additive table local model in XSPEC. From TCAF model spectral fits, 
physical parameters such as Keplerian disk rate, sub-Keplerian halo rate, shock location and 
compression ratio are extracted directly for better understanding of accretion processes around the 
BHC during this outburst. Low frequency quasi-periodic oscillation (QPO) are observed sporadically 
during the entire epoch of the outburst, with a general trend of increasing frequency during rising and 
decreasing frequency during declining phases of the outburst, as in other transient BHCs. The 
nature of the variation of the accretion rate ratio (ratio of  halo and disk rates) 
and QPOs (if observed), allows us to properly classify entire epoch of the outburst into following two spectral state, 
such as hard (HS), hard-intermediate (HIMS). These states are observed in the sequence of HS (Ris.) 
$\rightarrow$ HIMS (Ris.) $\rightarrow$ HIMS (Dec.) $\rightarrow$ HS (Dec.). 
This outburst of MAXI~J1836-194 could be termed as `failed' outburst, since no observation of 
soft (SS) and soft-intermediate (SIMS) spectral state are found during the entire outburst.
\end{abstract}

\keywords{Black hole:Individual (MAXI~J1836-194), accretion disk}

\bodymatter


\section{Introduction}
Most of the black holes (BHs) belong to binary systems with compact object as 
the primary star. Some of them are transient in nature. These binary black holes spend 
most of their lifetimes in quiescence states and exhibit occasional outbursts. During the 
outburst, they accrete matter from companion through Roche lobe or from winds.
As a result,  electromagnetic radiation is emitted from accretion disks in wide range, 
starting from radio to $\gamma$-rays. These transient black hole candidates (BHCs) are very 
interesting objects to study in X-rays, since they show rapid variation in their timing and 
spectral properties in X-radiation, which comes from the close vicinity of the BH and carries more 
information about it. One can find large number of scientific articles on observational and 
theoretical properties of these transient BHCs (\refcite{DD08,DD13,Nandi12} and references therein). 
Different spectral states, such as 
hard (HS), hard-intermediate (HIMS), soft-intermediate (SIMS), soft (SS), etc. are generally 
observed during an outburst of a transient BHC. It is also found that these spectral states 
form a hysteresis loop during an entire epoch of the outburst. Low frequency quasi-periodic 
oscillations (QPOs) could be observed in HS, HIMS and SIMS. These QPOs generally show monotonic 
evolutions (increasing/decreasing) during HS and HIMS of the rising/declining phases of the outburst, 
and during SIMS QPOs are observed sporadically on and off. Sometimes, these objects show high frequency 
QPOs in 3:2 ratios. The evolution of low frequency QPOs during rising and declining phases of the 
outbursts could be well explained with propagating oscillatory shock (POS) model 
\cite{C05,C08,DD10,DD13,Nandi12}.

There are several theoretical models available to describe the accretion disk dynamics. The emerging spectra 
from the disk basically consists of a thermal multi color black body component and a power law component due 
to thermal Comptonization. The black body component arises from the standard Keplerian disk \cite{SS73}
while power law component emerges from so-called $\it{`Compton}$ $\it{clouds}$' \cite{ST80,ST85}
located 'nearby'. The Compton cloud region is made of hot electrons.
Relatively cold soft photons coming from the standard disk, become hard after suffering multiple scattering 
with hot electrons at Compton cloud region.
In Two Component Advective Flow (TCAF) solution \cite{CT95,C97}
the so-called {\it'Compton clouds'} is replaced by CENtrifugal pressure supported BOundary Layer or CENBOL.
It is formed in the post-shock region.
The accreting materials of sub-Keplerian halo piles up at the centrifugal boundary to form a shock.
The other components with high viscosity and angular momentum submerges inside the halo components.
The Keplerian disk component emit soft photon which is re-emitted from the CENBOL as hard photon after repeated
scattering with hot electrons in CENBOL. 


Galactic BHC MAXI~J1836-194 was discovered by MAXI/GSC on 2011,August 29 \cite{Negoro11}.
It was simultaneously observed by Swift/BAT.
The optical observation determined its position at R.A. $= 18^h35^m43.43^s$, Dec $= -19^\circ 19'12.1''$ \cite{Kennea11}.
The BHC MAXI~J1836-194 is a highly rotating object with spin parameter $a=0.88 \pm 0.03$ \cite{Reis12}.
The orbital period is very short for this BHC, \textless $4.9 $ hrs \cite{Russell14}.
The inclination angle of the BHC is very low $4^\circ - 11^\circ $.
The mass and distance of the BHC are reported to be between $4-10$ $kPc$ and $4-12$ $M_{\odot}$ respectively \cite{Russell14}.
The companion is identified as Be star \cite{Cenko11}.
Strong jet is observed in radio and IR frequencies \cite{Trushkin11}.
Recently after the inclusion of TCAF model \cite{CT95} into HEASARC's spectral analysis software
package XSPEC as a local additive table model, accretion flow dynamics of few BHCs (GX~339-4, H~1743-322, MAXI~J1659-152)
are well understood (Ref. \refcite{DCM14,DMC15,DMCM15,MDC14})
This motivated us to study current outburst of MAXI~J1836-194 with the model. 

The paper is organized in the following way: in the next Section, we briefly discuss observation and data
analysis procedures using HEASARC's HeaSoft software package.
In \S 3, we present results of spectral analysis using TCAF solution based 
{\it fits} file as a local additive table model in XSPEC.
We also compare combined disk blackbody (DBB) and power-law (PL) model fitted spectral analysis results with those of the TCAF fitted analysis results. 
Finally, in \S 4, we present a brief discussion and make our concluding remarks.

\section{Observation and Data Analysis}

We analyze 2.5-25 keV RXTE/PCU2 archival data for 35 observation spread over the entire outburst.
RXTE started to observe the BHC after two days of its discovery on 2011, Aug. 31., (MJD = 55804).
We carry out data analysis using HEASARC's software package HeaSoft version HEADAS 6.15 and
XSPEC version 12.8.
For spectral analysis, we use TCAF model generated fits file in XSPEC to extract physical parameters
such as Keplerian disk rate ($\dot{m_d}$) and sub-Keplerian halo rate ($\dot{m_h}$) in 
Eddington rate, shock location ($X_s$) in Schwarzsachild radius and compression ratio (R).
We also fitted the data with combined DBB and PL model.
From combined DBB and PL model fitting, we get the information about
disk temperature ($T_{in}$), photon index($\Gamma$), DBB flux and PL flux.
From PDS, we get the information about the quasi periodic oscillation (QPO) frequency.

\begin{figure}[t]
\vskip -1.0cm
\centerline{
\includegraphics[scale=0.9,angle=0,width=12.0truecm]{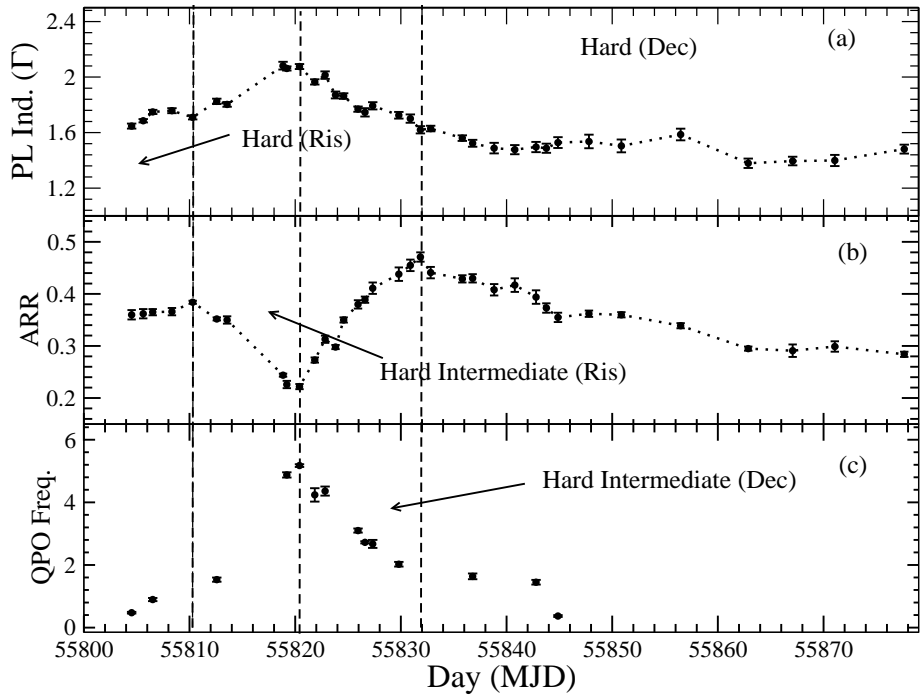}}
\caption{Variation of (a) Photon Index ($\Gamma$), (b) accretion rate ratio
(ARR; ratio between halo and disk rates), 
with day (MJD) for the 2011 outburst of MAXI J1836-194 are shown.
In the bottom panel (c), observed primary dominating QPO frequencies (in Hz) with day (MJD) are shown.
The vertical dashed lines indicate the transitions of between different spectral states.}
\label{fig1}
\end{figure}

\section{Results}

We study accretion flow dynamics of BHC MAXI J1836-194 during 2011 outburst with 2.5-25 keV RXTE/PCU2 data.
We fit the data with TCAF model fits file to extract physical parameter such as:
Keplerian disk rate ($\dot{m_d}$) in Eddington Rate, 
sub-Keplerian halo rate($\dot{m_h}$) in Eddington rate, 
Shock Location ($X_s$) in Schwarzschild radius and Compression Ratio ($R$).
We also fit the data with combined DBB and PL model to calculate
disk temperature ($T_{in}$), photon index ($\Gamma$), DBB flux and PL flux.
The variation of disk rate ($\dot{m_d}$), halo rate ($\dot{m_h}$), DBB flux, PL flux, 
total flux (DBB flux + PL flux) and total accretion flow ($\dot{m_d} + \dot{m_h}$) 
with day are shown in Ref.~\refcite{Jana16}. 
The variation of total flux and total accretion flow roughly matches.
The variation of DBB flux and PL flux also roughly match with the variation of 
disk rate ($\dot{m_d}$) and halo rate ($\dot{m_h}$) respectively.
In the Figure, we show the variation of Power law photon index, accretion rate ratio
$(ARR = \dot{m_h}/\dot{m_d} )$ and QPO frequency with day.
QPOs are observed sporadically on and off during the entire phase of the outburst.
The variation of the ARR and nature of QPOs (if observed), allows us to properly classify the entire 
epoch of the outburst into different spectral states, which are found to be in the following 
sequence: hard state (Rising) $\rightarrow$ hard-intermediate (Rising) $\rightarrow$ 
hard-intermediate (Declining) $\rightarrow$ hard state (Declining).

\noindent {i) \textbf{Hard State (Rising)}} - The BHC was in hard state for the first five observation with clear domination of 
halo rate $(\dot{m_{h}})$. Both the accretion rates, $\dot{m_d}$ and $\dot{m_h}$, increase and ARR becomes 
locally maximum on 2011 September 6 (MJD=55810.29) and the source enters in HIMS (Rising).
In the hard state of the rising phase, photon index was very low, varied  form 1.65 to 1.74.
We find QPOs only on 2 observations. On the first day of observation, on August 31, 2011 (MJD=55804.52), 
a QPO of frequency 0.47 Hz is observed. We find the other QPO of frequency 1.53 Hz on September 4, 2011 (MJD=55808.33).
We find QPO frequency in monotonically increasing nature which indicates that the oscillating shock may be moving 
inward, i.e., the Compton cloud is getting smaller.

\noindent {ii) \textbf{Hard Intermediate State (Rising)}} - The source was in HIMS for $\sim 10$ days.
Crucial $\sim 4$ days observation (MJD 55814-55817) is missing during this state.
We find QPOs on 3 observations.
On September 16 (MJD=55820.41), ARR reaches its minimum value with disk rate ($\dot{m_{d}}$) is maximum.
We find QPO of maximum frequency of 5.17 Hz on this day.
Photon Index ($\Gamma$) also attains maximum value ($\sim 2.08$) in this state.

\noindent {iii) \textbf{Hard Intermediate State (Declining)}} - The source enters on HIMS (declining phase) after
September 16 (MJD=55820.41).
The source stays at this state for next $\sim 11$ days.
QPOs are found sporadically.
QPO frequency is in monotonically decreasing nature,
it decreases from 5.17 Hz to 2.02 Hz.
The spectra started to become harder towards the end of the outburst.
Photon Index also decreases.
In this state sub-Keplerian halo rate ($\dot{m_h}$) starts increasing.
On September 27 (MJD=55831.85), ARR reaches its maximum value as it enters in Hard state in declining phase.

\noindent {iv) \textbf{Hard State (Declining)}} - The source was in the hard state till the end of the outburst.
We found the photon index around $\sim 1.4$ at the end of the outburst which is unusually low.
QPO was observed sporadically in 3 observations.
The shock was found to move away from the black hole.
The shock strength also increased in the hard state of the declining phase.

From the spectral fits, we also found that the TCAF model normalization comes out 
within a narrow range of $0.25-0.35$ except few days in HIMS when strong jet was 
present. Here, most importantly by keeping mass as a free parameter, we also found that 
for the best fitted spectra, the mass of the BHC is in the range of $7.5 M_{\odot}-11 M_{\odot}$.

\section{Discussion and Conclusion}

MAXI~J1836-184 is a strange object. It showed some unusual behavior during the outburst.
From the TCAF model fitted spectral evolution in conjunction with timing analysis, we 
have been successfully been able to understand accretion flow dynamics around the BHC 
MAXI~J1836-194 during its 2011 outburst. Two different spectral states (HS and HIMS) 
are classified, which form a hysteresis evolutions. Unlike other BHCs, SS and SIMS 
are absent during the entire outburst. This may be because, the black hole is 
immersed into an excretion disk of a high mass Be star, making the disk dominated by 
sub-Keplerian flow with low angular momentum.
In the rising phase, we found the trends of monotonically increasing nature of QPO frequency (0.47 Hz to 5.17 Hz).
In the declining phase, 
the QPO frequency shows a trend of monotonically decreasing nature (5.17 Hz to 0.37 Hz).
This indicates shock (which traces the size of the Compton cloud) moves closer to the black hole in the rising phase and
it moves away from the black hole in the declining phase.
We do not find QPOs in regular basis.
This may be due to non-satisfaction of resonance condition when 
infall time scale of the matter matches with cooling time scale at the shock location 
(Ref. \refcite{MCM15,C15}).
We found during the rising phase, halo rate $(\dot{m_h})$ attains maximum value on ~ MJD $\sim 55810$
while disk rate $(\dot{m_d})$ attains maximum after $\sim 10$ days on MJD $\sim 55820$.
Again in the declining phase of the outburst, halo rate $(\dot{m_h})$ attains the peak on MJD4 $\sim55826$,
while the disk rate $(\dot{m_d})$ attains the peak on MJD $\sim 55836$ after $\sim 10$ days 
(see, Figure 1 of Ref.~\refcite{Jana15}).
This indicates that the viscous time scale for the system is $\sim 10$ days.

\section*{Acknowledgments}
AJ and DD acknowledge support from ISRO sponsored RESPOND project fund (ISRO/RES/2/388/2014-15).
DD and DC acknowledge support from DST sponsored Fast-track Young Scientist project fund (SR/FTP/PS-188/2012).
AAM and SM acknowledge supports of MoES sponsored junior research fellowship and post-doctoral fellowship respectively.
We are also thankful to MG14 organizers, ICTP, and IUPAP for providing partial travel support and conference registration 
fee for the presenting author (DD).


\end{document}